\documentclass[aps,prd,twocolumn,showpacs,superscriptaddress,groupedaddress]{revtex4}  

\usepackage{dcolumn}   
\usepackage{graphicx}  
\usepackage{bm}        
\usepackage{amssymb,amsmath}   
\usepackage{epsfig}

\newcommand{\ppb}{\mbox{\ensuremath{p\bar p}}}

\newcommand{\invfb}{fb$^{-1}$}

\def\slashchar#1{\setbox0=\hbox{$#1$}           
  \dimen0=\wd0                                 
  \setbox1=\hbox{/} \dimen1=\wd1               
  \ifdim\dimen0>\dimen1                        
     \rlap{\hbox to \dimen0{\hfil/\hfil}}      
     #1                                        
  \else                                        
     \rlap{\hbox to \dimen1{\hfil$#1$\hfil}}   
     /                                         
  \fi}                                         %
\def\met{\slashchar{E}_T}                    %
\def\slashchar#1{\setbox0=\hbox{$#1$}           
  \dimen0=\wd0                                 
  \setbox1=\hbox{/} \dimen1=\wd1               
  \ifdim\dimen0>\dimen1                        
     \rlap{\hbox to \dimen0{\hfil/\hfil}}      
     #1                                        
  \else                                        
     \rlap{\hbox to \dimen1{\hfil$#1$\hfil}}   
     /                                         
  \fi}                                         %
\def\mpx{\slashchar{p}_x}                    %
\def\slashchar#1{\setbox0=\hbox{$#1$}           
  \dimen0=\wd0                                 
  \setbox1=\hbox{/} \dimen1=\wd1               
  \ifdim\dimen0>\dimen1                        
     \rlap{\hbox to \dimen0{\hfil/\hfil}}      
     #1                                        
  \else                                        
     \rlap{\hbox to \dimen1{\hfil$#1$\hfil}}   
     /                                         
  \fi}                                         %
\def\mpy{\slashchar{p}_y}                    %

\newcommand{\mht}{\mbox{\ensuremath{\slash\kern-.7emH_{T}}}}
\newcommand{\mpt}{\mbox{\ensuremath{\slash\kern-.5emp_{T}}}}

\newcommand{\SM}{SM}

\newcommand{\MLQ}{\ensuremath{M_{LQ}}}
\newcommand{\mlq}{\ensuremath{M_{LQ}}}
\newcommand{\LQbar}{\ensuremath{\overline{LQ}}}
\newcommand{\enuqq}{\ensuremath{eq \nu_{e} q'}}

\begin{document}

\hspace{5.2in} \mbox{FERMILAB-PUB-11/316-E}

\title{Search for first generation leptoquark pair production in the electron 
+ missing energy + jets final state}
%
\affiliation{Universidad de Buenos Aires, Buenos Aires, Argentina}
\affiliation{LAFEX, Centro Brasileiro de Pesquisas F{\'\i}sicas, Rio de Janeiro, Brazil}
\affiliation{Universidade do Estado do Rio de Janeiro, Rio de Janeiro, Brazil}
\affiliation{Universidade Federal do ABC, Santo Andr\'e, Brazil}
\affiliation{Instituto de F\'{\i}sica Te\'orica, Universidade Estadual Paulista, S\~ao Paulo, Brazil}
\affiliation{Simon Fraser University, Vancouver, British Columbia, and York University, Toronto, Ontario, Canada}
\affiliation{University of Science and Technology of China, Hefei, People's Republic of China}
\affiliation{Universidad de los Andes, Bogot\'{a}, Colombia}
\affiliation{Charles University, Faculty of Mathematics and Physics, Center for Particle Physics, Prague, Czech Republic}
\affiliation{Czech Technical University in Prague, Prague, Czech Republic}
\affiliation{Center for Particle Physics, Institute of Physics, Academy of Sciences of the Czech Republic, Prague, Czech Republic}
\affiliation{Universidad San Francisco de Quito, Quito, Ecuador}
\affiliation{LPC, Universit\'e Blaise Pascal, CNRS/IN2P3, Clermont, France}
\affiliation{LPSC, Universit\'e Joseph Fourier Grenoble 1, CNRS/IN2P3, Institut National Polytechnique de Grenoble, Grenoble, France}
\affiliation{CPPM, Aix-Marseille Universit\'e, CNRS/IN2P3, Marseille, France}
\affiliation{LAL, Universit\'e Paris-Sud, CNRS/IN2P3, Orsay, France}
\affiliation{LPNHE, Universit\'es Paris VI and VII, CNRS/IN2P3, Paris, France}
\affiliation{CEA, Irfu, SPP, Saclay, France}
\affiliation{IPHC, Universit\'e de Strasbourg, CNRS/IN2P3, Strasbourg, France}
\affiliation{IPNL, Universit\'e Lyon 1, CNRS/IN2P3, Villeurbanne, France and Universit\'e de Lyon, Lyon, France}
\affiliation{III. Physikalisches Institut A, RWTH Aachen University, Aachen, Germany}
\affiliation{Physikalisches Institut, Universit{\"a}t Freiburg, Freiburg, Germany}
\affiliation{II. Physikalisches Institut, Georg-August-Universit{\"a}t G\"ottingen, G\"ottingen, Germany}
\affiliation{Institut f{\"u}r Physik, Universit{\"a}t Mainz, Mainz, Germany}
\affiliation{Ludwig-Maximilians-Universit{\"a}t M{\"u}nchen, M{\"u}nchen, Germany}
\affiliation{Fachbereich Physik, Bergische Universit{\"a}t Wuppertal, Wuppertal, Germany}
\affiliation{Panjab University, Chandigarh, India}
\affiliation{Delhi University, Delhi, India}
\affiliation{Tata Institute of Fundamental Research, Mumbai, India}
\affiliation{University College Dublin, Dublin, Ireland}
\affiliation{Korea Detector Laboratory, Korea University, Seoul, Korea}
\affiliation{CINVESTAV, Mexico City, Mexico}
\affiliation{Nikhef, Science Park, Amsterdam, the Netherlands}
\affiliation{Radboud University Nijmegen, Nijmegen, the Netherlands and Nikhef, Science Park, Amsterdam, the Netherlands}
\affiliation{Joint Institute for Nuclear Research, Dubna, Russia}
\affiliation{Institute for Theoretical and Experimental Physics, Moscow, Russia}
\affiliation{Moscow State University, Moscow, Russia}
\affiliation{Institute for High Energy Physics, Protvino, Russia}
\affiliation{Petersburg Nuclear Physics Institute, St. Petersburg, Russia}
\affiliation{Instituci\'{o} Catalana de Recerca i Estudis Avan\c{c}ats (ICREA) and Institut de F\'{i}sica d'Altes Energies (IFAE), Barcelona, Spain}
\affiliation{Stockholm University, Stockholm and Uppsala University, Uppsala, Sweden}
\affiliation{Lancaster University, Lancaster LA1 4YB, United Kingdom}
\affiliation{Imperial College London, London SW7 2AZ, United Kingdom}
\affiliation{The University of Manchester, Manchester M13 9PL, United Kingdom}
\affiliation{University of Arizona, Tucson, Arizona 85721, USA}
\affiliation{University of California Riverside, Riverside, California 92521, USA}
\affiliation{Florida State University, Tallahassee, Florida 32306, USA}
\affiliation{Fermi National Accelerator Laboratory, Batavia, Illinois 60510, USA}
\affiliation{University of Illinois at Chicago, Chicago, Illinois 60607, USA}
\affiliation{Northern Illinois University, DeKalb, Illinois 60115, USA}
\affiliation{Northwestern University, Evanston, Illinois 60208, USA}
\affiliation{Indiana University, Bloomington, Indiana 47405, USA}
\affiliation{Purdue University Calumet, Hammond, Indiana 46323, USA}
\affiliation{University of Notre Dame, Notre Dame, Indiana 46556, USA}
\affiliation{Iowa State University, Ames, Iowa 50011, USA}
\affiliation{University of Kansas, Lawrence, Kansas 66045, USA}
\affiliation{Kansas State University, Manhattan, Kansas 66506, USA}
\affiliation{Louisiana Tech University, Ruston, Louisiana 71272, USA}
\affiliation{Boston University, Boston, Massachusetts 02215, USA}
\affiliation{Northeastern University, Boston, Massachusetts 02115, USA}
\affiliation{University of Michigan, Ann Arbor, Michigan 48109, USA}
\affiliation{Michigan State University, East Lansing, Michigan 48824, USA}
\affiliation{University of Mississippi, University, Mississippi 38677, USA}
\affiliation{University of Nebraska, Lincoln, Nebraska 68588, USA}
\affiliation{Rutgers University, Piscataway, New Jersey 08855, USA}
\affiliation{Princeton University, Princeton, New Jersey 08544, USA}
\affiliation{State University of New York, Buffalo, New York 14260, USA}
\affiliation{Columbia University, New York, New York 10027, USA}
\affiliation{University of Rochester, Rochester, New York 14627, USA}
\affiliation{State University of New York, Stony Brook, New York 11794, USA}
\affiliation{Brookhaven National Laboratory, Upton, New York 11973, USA}
\affiliation{Langston University, Langston, Oklahoma 73050, USA}
\affiliation{University of Oklahoma, Norman, Oklahoma 73019, USA}
\affiliation{Oklahoma State University, Stillwater, Oklahoma 74078, USA}
\affiliation{Brown University, Providence, Rhode Island 02912, USA}
\affiliation{University of Texas, Arlington, Texas 76019, USA}
\affiliation{Southern Methodist University, Dallas, Texas 75275, USA}
\affiliation{Rice University, Houston, Texas 77005, USA}
\affiliation{University of Virginia, Charlottesville, Virginia 22901, USA}
\affiliation{University of Washington, Seattle, Washington 98195, USA}
\author{V.M.~Abazov} \affiliation{Joint Institute for Nuclear Research, Dubna, Russia}
\author{B.~Abbott} \affiliation{University of Oklahoma, Norman, Oklahoma 73019, USA}
\author{B.S.~Acharya} \affiliation{Tata Institute of Fundamental Research, Mumbai, India}
\author{M.~Adams} \affiliation{University of Illinois at Chicago, Chicago, Illinois 60607, USA}
\author{T.~Adams} \affiliation{Florida State University, Tallahassee, Florida 32306, USA}
\author{G.D.~Alexeev} \affiliation{Joint Institute for Nuclear Research, Dubna, Russia}
\author{G.~Alkhazov} \affiliation{Petersburg Nuclear Physics Institute, St. Petersburg, Russia}
\author{A.~Alton$^{a}$} \affiliation{University of Michigan, Ann Arbor, Michigan 48109, USA}
\author{G.~Alverson} \affiliation{Northeastern University, Boston, Massachusetts 02115, USA}
\author{G.A.~Alves} \affiliation{LAFEX, Centro Brasileiro de Pesquisas F{\'\i}sicas, Rio de Janeiro, Brazil}
% ...
%
%
\collaboration{The D0 Collaboration\footnote{with visitors from
$^{a}$Augustana College, Sioux Falls, SD, USA,
$^{b}$The University of Liverpool, Liverpool, UK,
$^{c}$SLAC, Menlo Park, CA, USA,
$^{d}$University College London, London, UK,
$^{e}$Centro de Investigacion en Computacion - IPN, Mexico City, Mexico,
$^{f}$ECFM, Universidad Autonoma de Sinaloa, Culiac\'an, Mexico,
and 
$^{g}$Universit{\"a}t Bern, Bern, Switzerland.
$^{\ddag}$Deceased.
}} \noaffiliation
\vskip 0.25cm

\date{July 10, 2011}

\begin{abstract}
We present a search for the pair production of first generation 
scalar leptoquarks ($LQ$) in data corresponding to an integrated luminosity of 
5.4~\invfb{} ~collected with the D0 detector  
at the  Fermilab Tevatron Collider 
in \ppb\ collisions at $\sqrt{s}=1.96$\,TeV.
In the channel $LQ \LQbar \rightarrow \enuqq$, where $q,q'$ are $u$ or $d$ quarks, no significant excess of data 
over background is observed, and we set a 95\% C.L. lower limit of 326 GeV 
on the leptoquark mass, assuming equal
probabilities of leptoquark decays to $eq$ and $\nu_e q'$.
\end{abstract}

\pacs{13.85.Rm, 14.80.Sv}
\maketitle

Because of the limitations of the standard model
(SM), several extensions have been proposed,
among them supersymmetry (SUSY), 
grand unified theories, and string theory. Many of these extensions predict the existence 
of particles 
that directly connect the lepton and quark sectors.
By combining leptons and quarks in multiplets of a larger 
symmetry group, they are expected to interact through new mediating
bosons called leptoquarks ($LQ$)~\cite{Acosta:1999ws,Nakamura:2010zzi}. Leptoquarks 
can be either scalar or vector fields. This Letter will focus on the search
for scalar leptoquarks, and in the following we will
not distinguish particles from antiparticles.
Because effective models are assumed for experimental searches for
leptoquarks, our search is independent of specific extensions of the \SM{}.

In $\ppb$ collisions such as at the Tevatron Collider, leptoquarks can be produced 
in leptoquark-antileptoquark pairs. 
Leptoquark pair production can occur via both quark-antiquark 
annihilation and gluon-gluon fusion,
although quark-antiquark annihilation is dominant. 
The production cross section for scalar leptoquarks depends only on the strong coupling constant and on the leptoquark mass, and is known 
at next-to-leading~order (NLO)~\cite{Kramer:1997hh}.

Once produced, leptoquarks can decay to two final states: $lq$ and $\nu q'$ 
(where $l = e,~\mu,~\mbox{ or }~\tau$). 
It is assumed that in the low energy limit there is no intergenerational 
mixing. 
For first generation
$LQ$ pairs the final state will contain a pair of leptons
($e$ or $\nu_e$) and a pair of quarks ($u$ or $d$).
In this Letter, the case in which
one leptoquark decays to  $eq$ and the other to 
$\nu_{e} q'$ is considered (charge conjugate states are
assumed in the Letter).

We define $\beta$ to be the branching ratio of a first generation
leptoquark to decay to $eq$.
Then the probability for a leptoquark to decay to $\nu_e q'$ is $(1-\beta)$, 
and the probability for a leptoquark pair to decay to the final state \enuqq{} is
 $BR(LQ\LQbar\rightarrow \enuqq) = 2\beta(1-\beta)$.
Thus, the probability for the final state \enuqq{} is maximized when $\beta = 0.5$. 

Limits on the production of first generation leptoquarks have been reported
by the DELPHI~\cite{Abreu:1998fw}, OPAL~\cite{Abbiendi:2003iv,Abbiendi:2001sw}, H1~\cite{Aktas:2005pr}, ZEUS~\cite{Chekanov:2003af}, CDF~\cite{Acosta:2005ge}, and D0~\cite{Abazov:2009gf} Collaborations. 
Recently, CMS~\cite{Khachatryan:2010mp,Chatrchyan:2011ar}, and
ATLAS~\cite{Collaboration:2011uv} published the first searches for
scalar leptoquark pair production at the CERN LHC.

The D0 detector consists of tracking, calorimeter, and muon systems
~\cite{Abazov:2005pn,Abolins:2007yz,Angstadt:2009ie}. 
The central-tracking system consists of a silicon microstrip tracker and a central fiber tracker, 
both located within a 2~T superconducting solenoid.
A liquid-argon and uranium calorimeter consists of a central section (pseudorapidity $|\eta|< 1.1$~\cite{eta}) and two end sections ($1.5 < |\eta| <  4.2$).
The calorimeters have fine transverse and longitudinal segmentation with three 
principal layers identified as electromagnetic, and fine and 
coarse hadronic.
An outer muon system ($|\eta|<2$) 
consists of a layer of tracking detectors and scintillation trigger 
counters in front of 1.8~T toroids, followed by two similar layers 
after the toroids~\cite{Abazov:2005uk}. 
Data were collected with the D0 detector at the Fermilab Tevatron $p\bar{p}$
Collider operating at $\sqrt{s} = 1.96$~TeV between August 2002  and June 
2009, and correspond to an integrated luminosity of 5.4~\invfb.

An electron is identified from energy deposits in the electromagnetic 
calorimeter that are consistent with the shower development
expected for an electron and have a
matching track extrapolated from the central tracker.

Jets are reconstructed using a midpoint cone algorithm, 
with a cone size 
of 0.5~\cite{Baur:2000bi}. 
The jet energy  is corrected to the particle level using 
jet energy scale corrections determined from data~\cite{Abazov:2008hua}. 
The missing transverse energy ($\met$) is reconstructed from all the cells 
of the electromagnetic and hadronic calorimeters, except for the 
coarse hadronic sector where a noise-reduction algorithm is applied.  
Additional corrections are then applied for all identified 
objects including jets, electrons, and muons.

Events must satisfy at least one trigger from the single-electron and 
electron+jets suites of triggers.  For all data samples, trigger objects 
are required to match the reconstructed objects.
The trigger efficiencies are measured in data and parameterized 
for specific lepton and jet identification criteria. 


Scalar leptoquark pair Monte Carlo (MC) samples are generated 
using {\sc pythia}~\cite{Sjostrand:2006za} with 
{\sc CTEQ6L1}~\cite{Pumplin:2002vw} parton density
 functions. 
Signal samples are produced for different $LQ$ masses between 200 and 360~GeV. 
The corresponding cross sections at NLO are listed in 
Table~\ref{tab:lq_cross_section}.

\begin{table}
\caption{\label{tab:lq_cross_section}Scalar $LQ$ pair production cross sections, calculated at NLO, for different $\MLQ$~\cite{Kramer:1997hh}.}
\begin{ruledtabular}
  \begin{tabular}{ c | ccccccccccccccc  }
$\mlq$ (GeV) & 200 & 210 & 220 & 230 & 240 & 250 & 260 & 270 \\
\hline
 $\sigma$ (fb) & 268 & 193 & 141 & 103 & 76 & 56 & 42 & 31 \\
\hline \hline
$\mlq$ (GeV) &  280 & 290 & 300 & 310 & 320 & 340 & 360\\
\hline
 $\sigma$ (fb) & 23 & 17 & 13 & 10 & 7.4 & 4.2 & 2.4
  \end{tabular}
\end{ruledtabular}
\end{table}

Diboson ($WW$, $WZ$ and $ZZ$) background samples are produced with {\sc pythia} making use of the 
parton distribution functions {\sc CTEQ6L1}. The $t\bar{t}$ and $V (V=W$ or $Z)+$jets events 
are simulated with the matrix-element generator {\sc alpgen}~\cite{Mangano:2002ea}, 
interfaced to {\sc pythia} for subsequent parton showering and hadronization.  
Single top quark production is simulated
using {\sc comphep}~\cite{Boos:2004kh}. 
The cross sections for background processes are calculated at 
NLO (diboson~\cite{Campbell:1999ah}) and next-to-next-to-leading~order (NNLO)
($V$+jets~\cite{Campbell:2002tg} and $t\bar{t}$~\cite{Moch:2008qy}).  
We correct the generated spectrum of the transverse momentum ($p_T$) of the $Z$ boson in MC to match a 
corresponding dedicated measurement~\cite{Abazov:2007nt}. The $p_T$ 
spectrum of the $W$ boson is corrected taking into account the differences between predicted $Z$ and $W$ boson $p_T$ 
spectra at NNLO~\cite{Melnikov:2006kv}.

A full {\sc geant}-based detector simulation program~\cite{geant}, followed by the same reconstruction program as utilized for data, is used to 
process signal and background events from MC. In order to
model detector noise and contributions from the presence of additional $\ppb$ interactions,  events from randomly selected
beam crossings with the same instantaneous luminosity
profile as data are overlaid on the simulated events. 
Background from multijet production (MJ), where one 
of the jets mimics an electron, is evaluated from data using 
a data driven technique~\cite{Abazov:2008kt}.
In MC simulations, electron energies are corrected so they match those obtained from data. In addition, residual differences in jet
energy scale and resolution between data and MC are
reduced by applying dedicated corrections to MC events.

In the \enuqq{} final state, it is not known \emph{a priori} how to assign
the jets to the leptoquark decaying to $eq$ or $\nu_e q'$.
Therefore, to reconstruct the properties such as mass and $p_T$ of the 
leptoquarks from the final products, an algorithm  is needed
to choose 
the best pairing. We do not impose a requirement on the number
of jets, but we use only the two leading $p_T$ jets for
pairings. There are two possible combinations, corresponding to the
leading jet pairing with either the electron or the neutrino. We found
that it is most effective to choose the pairing that minimizes
the difference between the transverse masses, $M_T=\sqrt{E_T^2-\vec{p}_T^2}$, where $E_T$ and $\vec{p}_T$ are the transverse energy 
and the transverse momentum vector of the two $LQ$s.
This pairing algorithm is successful 
in making the correct assignment in about 75\% of MC signal events.

Events are selected to be consistent with the $LQ \LQbar \rightarrow \enuqq$ process. We require one electron with $p_T > 15$~GeV in the central calorimeter 
region $|\eta_e|<1.1$; $\met >15$~GeV, to be consistent with the undetected neutrino; and at least two jets with $p_T >$ 20 GeV and $|\eta_{\text{jet}}|<2.5$.
To suppress MJ background, events 
are required to satisfy
$\met/50 + M_{T}^{e\nu}/70 \geq 1$, where $M_{T}^{e\nu}$ is the transverse mass of the $(e,\nu)$ combination, and $\met$ and $M_{T}^{e\nu}$ are in GeV.

\begin{figure}
\includegraphics[width=7cm]{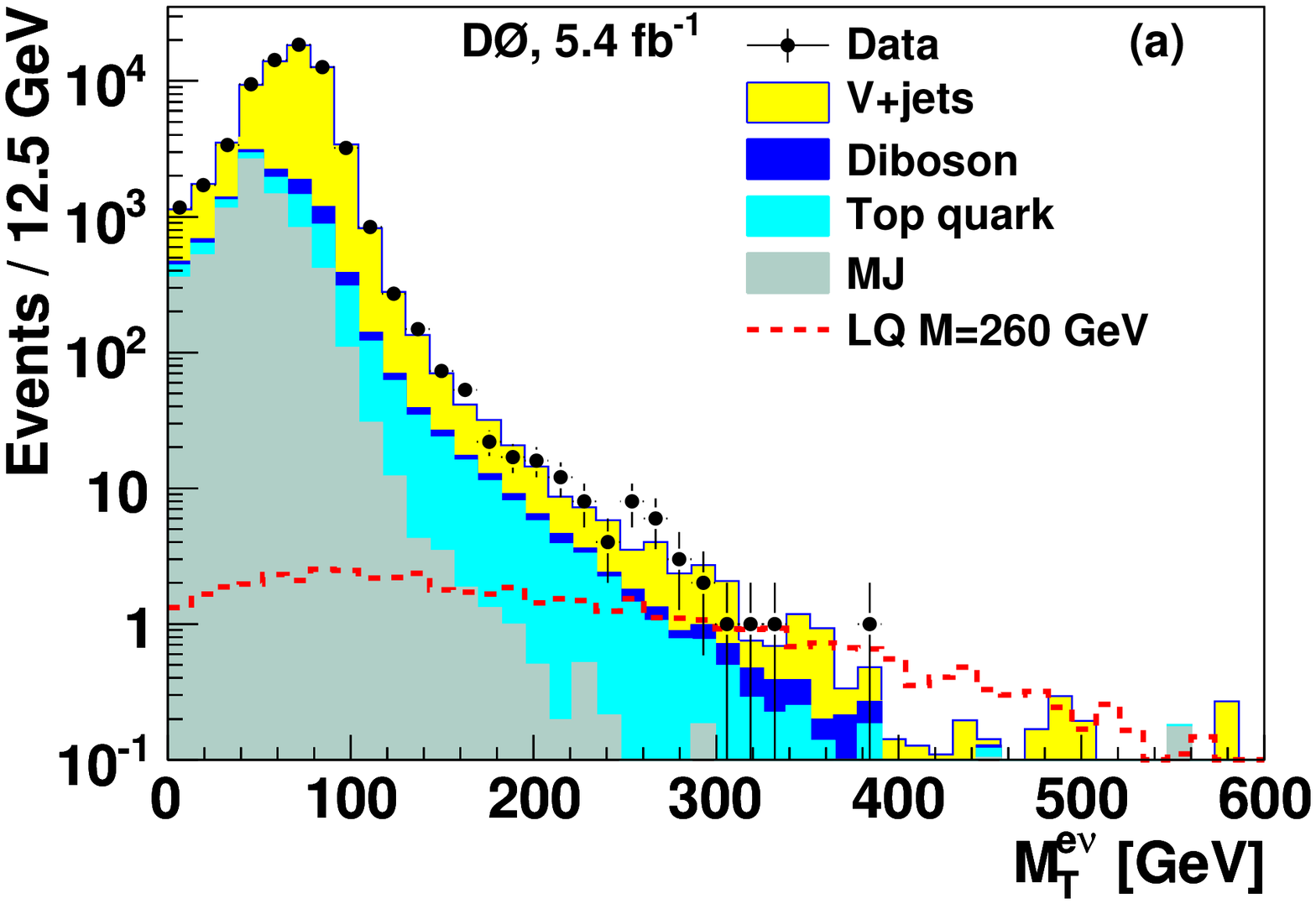} 
\vspace{-0.2cm}
\includegraphics[width=7cm]{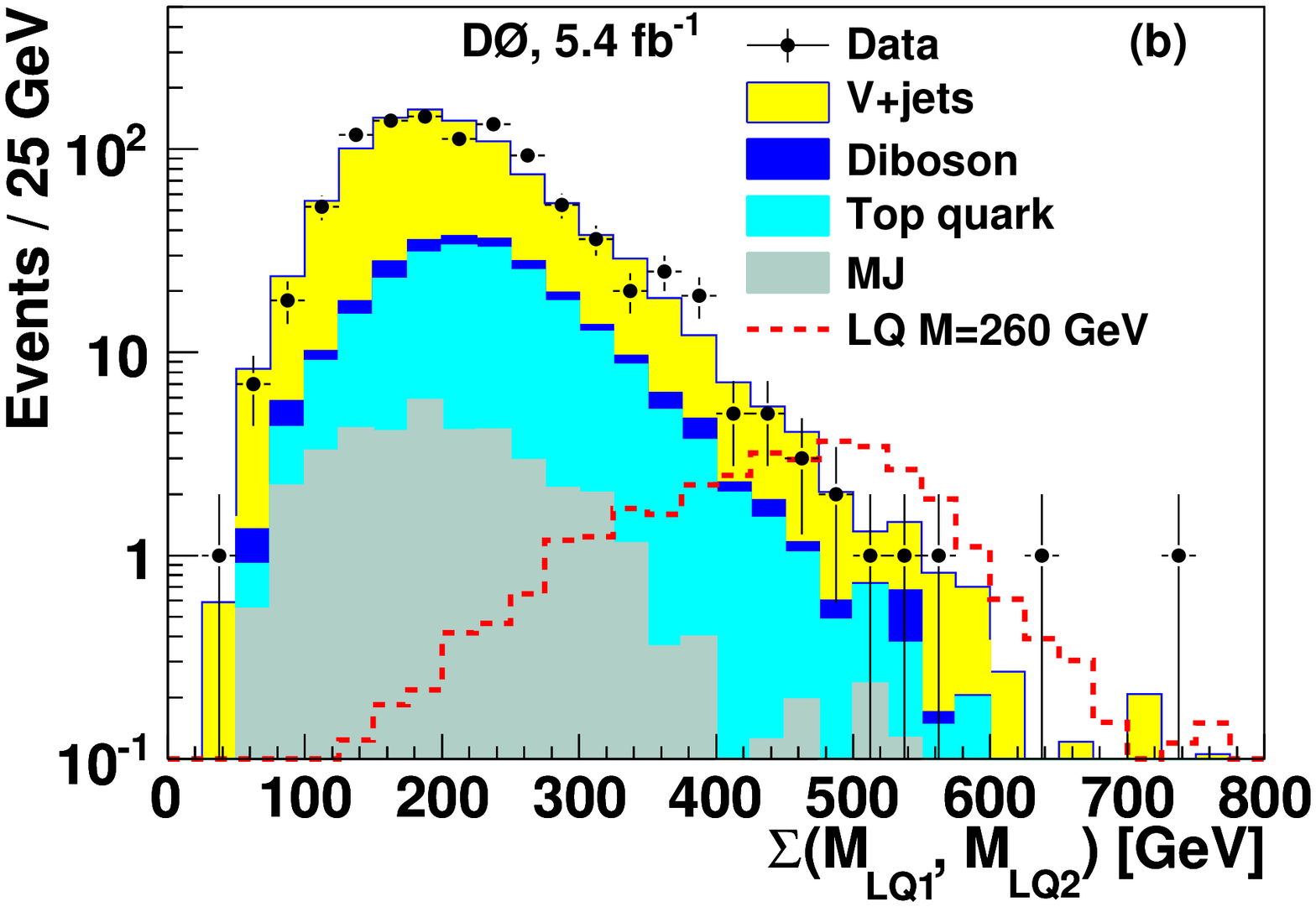}
\vspace{-0.2cm}
\includegraphics[width=7cm]{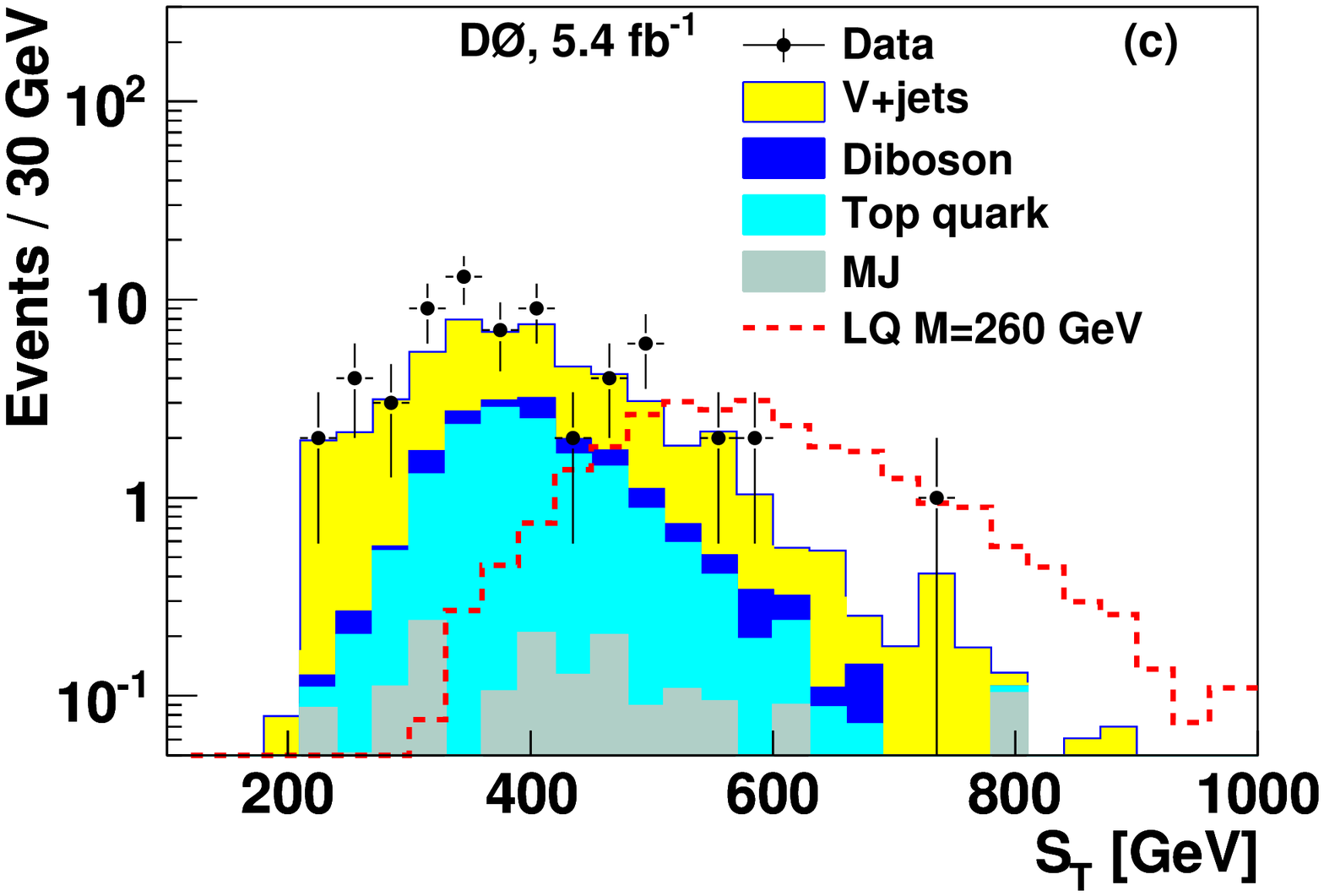}
\vspace{-0.2cm}
\includegraphics[width=7cm]{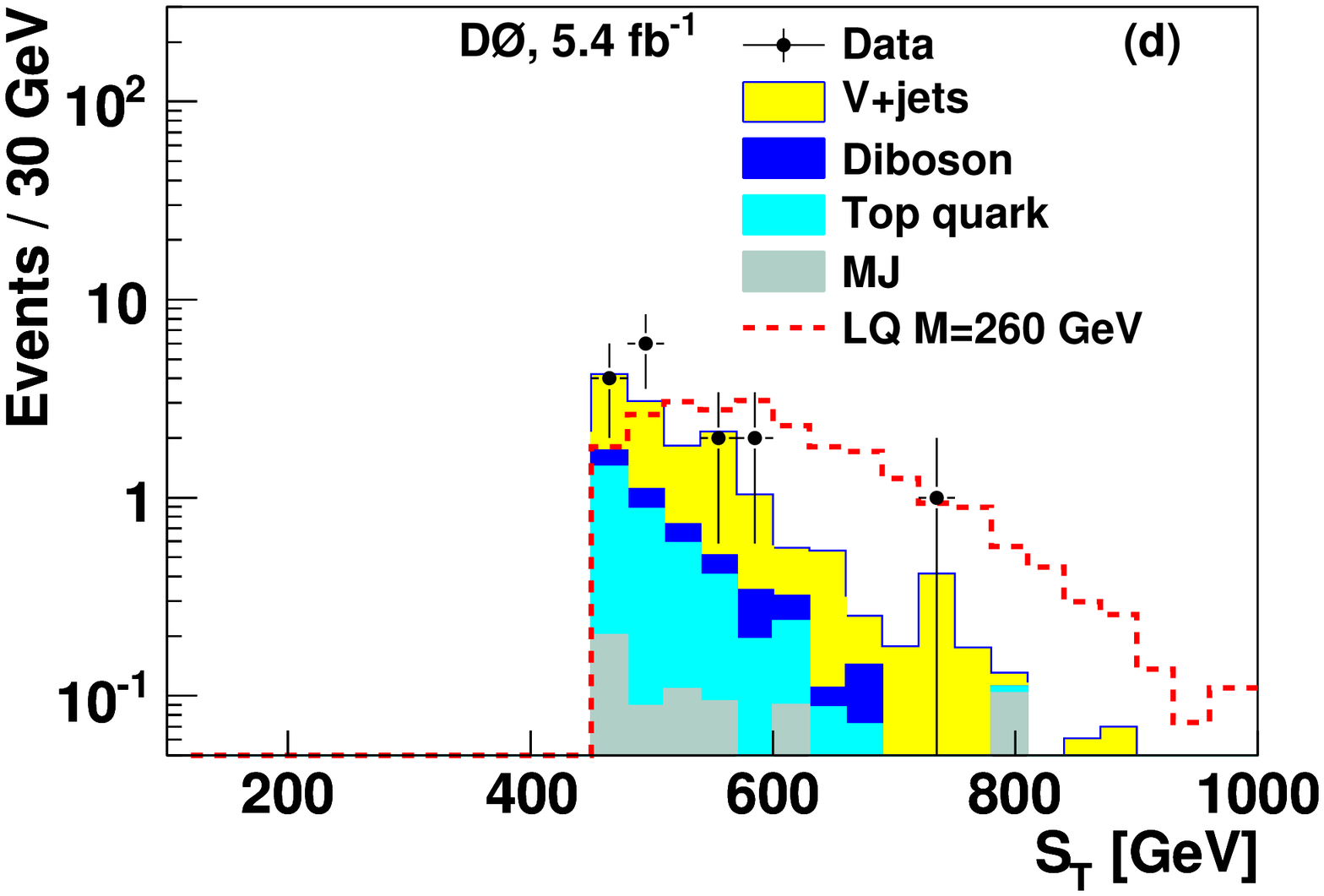}
\vspace{-0.2cm}
\caption{\label{fig:dist} (color online) (a) $M_{T}^{e\nu}$ distribution after preselection,
  (b) $\sum{\mlq}$ for $M_{T}^{e\nu}>110$~GeV, (c)
  the $S_T$ for $M_{T}^{e\nu}>110$~GeV and $\sum{\mlq}>350$~GeV, and (d) the $S_T$
  distribution after the final selection, which is used to set an upper limit on the $LQ$ pair production
  cross section.}

\end{figure}

At this stage we observe 65992 data events, while we expect
$65703\pm 61(\rm stat)\pm 5958(sys)$ from SM background and
$50.4\pm 0.4(\rm stat)\pm 6.8(sys)$ events from scalar $LQ$
production for $\MLQ=260$~GeV and
$\beta=0.5$. Figure~\ref{fig:dist}(a) shows the $M_{T}^{e\nu}$ distribution for the data and \SM{}
processes.
Data are consistent with the SM
predictions. To reduce the dominant SM $V$+jets background, we require 
$M_{T}^{e\nu} \geq 110$ GeV.  
The pairing algorithm described previously allows us to reconstruct $\MLQ$. 
Since the longitudinal component of the neutrino
momentum, $p_z$, is not measurable, we reconstruct  
only the visible mass of the decay $LQ\rightarrow \nu_e q'$ as
$\mlq=M(\text{jet}+\nu_{\text{vis}})$, where the four vector of
$\nu_{\text{vis}}$ is given as $(\mpx, \mpy, 0,
\met)$. Figure~\ref{fig:dist}(b) shows the distribution of the sum $\sum{\mlq}$ of the invariant mass
of the decay $LQ\rightarrow e q$ and the visible mass of the 
decay $LQ\rightarrow \nu_e q'$ after the requirement $M_{T}^{e\nu}\geq 110$~GeV. We then use
$\sum{\mlq}$ to reduce \SM{} backgrounds, further requiring that $\sum{\mlq}>350$~GeV. Finally, we require that the scalar sum
of the $p_T$ of the lepton, the $\met$, and the two jets, $S_T$,
shown in Fig.~\ref{fig:dist}(c) after all selections, be greater than
450~GeV.  Selection criteria are optimized to achieve the 
best expected sensitivity  for $\MLQ=260$~GeV. This yields 15 observed events
for an expected background of $14.8\pm 0.6(\rm stat)\pm 1.1(sys)$
events.  The event counts after each requirement are shown
in Table \ref{tab:event_count_after_cuts}.

\begin{table}
 \caption{\label{tab:event_count_after_cuts}Event counts and the predicted number of signal events for $\MLQ=260$~GeV and $\beta=0.5$ after each selection requirement.}
\begin{ruledtabular}
 \begin{tabular}{r | c  r@{$\,\pm \,$}l r@{$\,\pm \,$}l}

 & Data  & \multicolumn{2}{c}{Total background} &  \multicolumn{2}{c}{Signal} \\
  \hline{}

  Preselection        & 65992 & 65703 & 5958  &  50 &  7 \\
 $M_{T}^{e\nu} > 110$ GeV & 990 & 986 & 82  &  34 & 5  \\
  $\sum{\MLQ} > 350 $ GeV & 64 & 55& 4  &  27 & 4  \\
  $S_{T} > 450 $ GeV         & 15 & 15 &  1  &   24 & 3  \\

  \end{tabular}
\end{ruledtabular} 
\end{table}

Systematic uncertainties which affect only the
normalization of the background and the signal efficiency include uncertainties on cross sections of signal (10\%) and 
background ($6\%-10\%$) processes, normalization of the MJ background (20\%),
integrated luminosity (6.1\%), and lepton trigger and identification (4\%). 
Uncertainties which also affect the differential distribution
of $S_T$ which is the quantity used to set the limits on $LQ$ are due to the jet energy resolution and scale, jet identification
efficiency, parton distribution functions, and the modeling of the jet
$p_T$ distribution of the dominant $W$+jets background.  
Their impacts are evaluated by repeating the
analysis with values varied by $\pm 1$~standard deviation (SD).  For the uncertainty on the jet
$p_T$ modeling, the impact is estimated by comparing the
jet $p_T$ distributions 
between {\sc alpgen} and data unfolded to particle level from the recent D0
measurement~\cite{Abazov:2011rf}. The ratio is applied as weight to the $W$+jets jet $p_T$ distribution, 
and the new distribution is taken as $\pm 1$~SD band.

\begin{figure}[htbp]
\centering
\includegraphics[width=8cm]{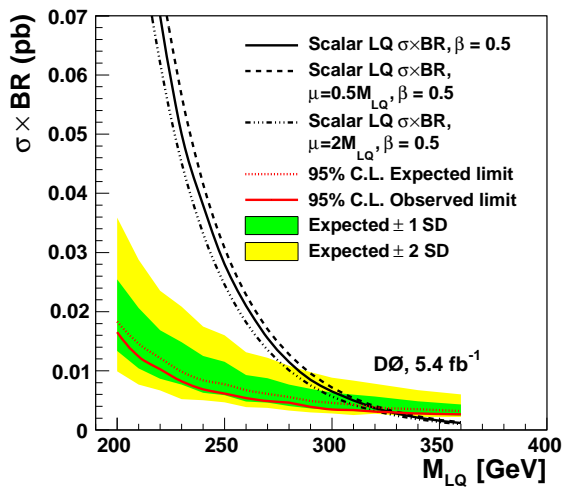}
\caption {(color online) Expected and observed upper limits calculated at the $95\%$
  C.L. on the $LQ$ cross section as a function of $\MLQ$ for a scalar
  leptoquark compared with the NLO prediction for $\beta=0.5$. The NLO
  cross section is shown for different choices of the renormalization
  and factorization scales, $\mu=M_{LQ}$, $\mu=0.5\times M_{LQ}$, and $\mu=2\times M_{LQ}$. }

\label{fig:limit}
\end{figure}

\begin{figure}[htbp]
\centering
\includegraphics[width=8cm]{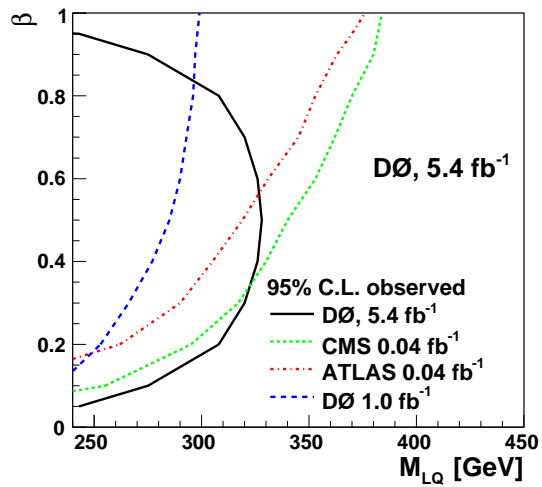}
\caption{\label{fig:limit2d} (color online) $95\%$ C.L. observed limit  for $\mu=M_{LQ}$ 
on the
  leptoquark mass as a function of $\beta$ compared
  with the previous D0 result~\cite{Abazov:2009gf}, and CMS~\cite{Khachatryan:2010mp,Chatrchyan:2011ar} and
   ATLAS~\cite{Collaboration:2011uv} results.  }
\end{figure}

The distribution of the $S_{T}$ after all selection requirements, shown in Fig.~\ref{fig:dist}(d), is used as a discriminant 
to set an upper limit on the $LQ$ pair production cross section
in the \enuqq{} channel.
For each generated $\MLQ$, the limit is calculated at the 95\% C.L. using the
semi-frequentist $CL_s$ method based on a Poisson log-likelihood test 
statistic~\cite{Junk:1999kv}.   
Signal and background normalizations 
and shape variations due to systematic uncertainties are incorporated 
assuming Gaussian priors.
The best fit to the 
background distributions is evaluated by minimizing a profile likelihood 
function with respect to the observed data and various
sources of uncertainty, maintaining all correlations among systematic 
uncertainties~\cite{wadePublic}. 
Limits on the cross section multiplied by the branching fraction and
the theoretical $LQ$ cross section for $\beta=0.5$ are shown in Fig.~\ref{fig:limit}. 
The limit on the $LQ$ mass as a function of $\beta$ is
determined as shown in Fig.~\ref{fig:limit2d},
 and compared to the previous D0~\cite{Abazov:2009gf}, CMS~\cite{Khachatryan:2010mp,Chatrchyan:2011ar},  and 
 ATLAS \cite{Collaboration:2011uv} results.

In summary, we have searched for scalar leptoquark pair production in the \enuqq{}
final state in 5.4 fb$^{-1}$ of integrated luminosity of $p\bar{p}$ collisions
at $\sqrt{s}=1.96$~TeV. In the absence of a signal, we exclude the production of first generation leptoquarks with 
$\MLQ<326$~GeV for $\beta=0.5$ at the $95\%$ C.L.

%
We thank the staffs at Fermilab and collaborating institutions,
and acknowledge support from the
DOE and NSF (USA);
CEA and CNRS/IN2P3 (France);
FASI, Rosatom and RFBR (Russia);
CNPq, FAPERJ, FAPESP and FUNDUNESP (Brazil);
DAE and DST (India);
Colciencias (Colombia);
CONACyT (Mexico);
KRF and KOSEF (Korea);
CONICET and UBACyT (Argentina);
FOM (The Netherlands);
STFC and the Royal Society (United Kingdom);
MSMT and GACR (Czech Republic);
CRC Program and NSERC (Canada);
BMBF and DFG (Germany);
SFI (Ireland);
The Swedish Research Council (Sweden);
and
CAS and CNSF (China).
%

\bibliographystyle{h-physrev3}
\bibliography{d0LQ1genRef}

\begin{thebibliography}{10}
\bibitem{eta}
The pseudorapidity is defined as $\eta=-\ln[\tan(\theta/2)]$, where $\theta$ is
  the polar angle with respect to the proton beam direction.
\end{thebibliography}

\end{document}